\documentstyle[aps]{revtex}
\begin{document}
%      Title page and abstract...
\begin{title}
{\Large\bf Solvable three boson model with attractive
delta function interactions}
\end{title}
\author{\large J. G. Muga\footnote{Permanent address:
Departamento de F\'{\i}sica Fundamental
y Experimental,  Universidad de La Laguna,
La Laguna, Tenerife, Spain} and R. F. Snider}
\address{Department of Chemistry, University of British Columbia,
Vancouver, B.C., Canada V6T 1Z1}

\date{\today}
\maketitle

\begin{abstract}
A one parameter solvable model for three bosons subject to delta
function attractive interactions in one-dimension with periodic
boundary conditions is studied. The energy levels and wave functions
are classified and given explicitly in terms of three momenta. In
particular, eigenstates and eigenvalues are described as functions of
the model parameter, $c$. Some of the states are given in terms of
complex momenta and represent dimer or trimer configurations for large
negative $c$. The asymptotic behaviour for small and large values of
the parameter, and at thresholds between real and complex momenta is
provided.  The properties of the potential energy are also discussed.

\end{abstract}


\section{Introduction}

        Quantum solvable models where the wave functions, energy
eigenvalues and other quantities of physical interest can be obtained
explicitly in terms of known functions or with minimum numerical
effort (typically by solving a transcendental equation or by
quadratures) are useful to test and refine concepts and/or numerical
methods, and as first approximations to more realistic systems.
Occasionally unexpected physical phenomena are revealed \cite{Lieb97}.
In this paper we shall analyze a one parameter model for three bosons
subject to attractive delta function pair interactions in
one-dimension with periodic boundary conditions (contrast this to
three particles ``on a ring'', see \cite{McG97}).  Our original
motivation was to examine a system with attractive forces where single
and compound particles may coexist. This is of particular interest
when studying the kinetic theory of gases composed by particles that
can form stable aggregates (such as dimers or trimers), especially when
chemical reactions can occur \cite{Lowry Snider}. This article deals
exclusively with the model itself, which has been found to be quite
complex in mathematical detail.

        The literature on one-dimensional solvable models of three,
and generally $N$, particles is rather extensive. These models could
be primarily classified according to the type of interaction involved
\cite{rev}. However, even with the same interaction but with different
boundary conditions, different formal treatments are required and very
different results may be found.  Periodic boundary conditions are
suitable for modelling a gas or a crystal lattice in the thermodynamic
limit.  In contrast, in the limit as the box length becomes large,
information about the corresponding scattering problem of a
one-dimensional system can be extracted \cite{Kosloff,Muga-Snider}.
Indeed these are the standard boundary conditions for actual
calculations of time dependent wave function scattering \cite{TD}. The
model studied here is a particular case of the ``interacting bose
gas'' of Lieb and Liniger \cite{LL}, who examined $N$ particles
subject to 2-body delta function interactions and boson symmetry.
Further analysis of this gas was carried out in several papers
\cite{L,Yang67,Sut68,Yang69}, but, having different objectives and
applications in mind, in none of these works was the attractive case
examined, except for the appendix on the $N=2$ case in \cite{LL}.
Lieb and Liniger found some unexpected effects of the periodic
confinement but did not investigate the analogous effects for $N>2$.
In a series of papers \cite{McGf1,McGf2,McGf3}, McGuire has examined a
related one-dimensional many-particle Fermion system with one particle
having spin down in a sea of spin up particles, interacting via
$\delta$-function potentials (both repulsive and attractive).  For
other models with periodic boundary conditions but different
interactions, see \cite{Sut71,Sut71c,Sut72,CR,Quesne}.

        If the particles are not confined in a box, the wave function
obeys the standard vanishing boundary conditions (for bound states) or
scattering boundary conditions, at infinite distances \cite{Newton}.
However, in most available models, rearrangement processes where a
bound pair collides with a single particle to form a new pair are not
allowed and cannot be examined. An exception is the work of McGuire on
the attractive, two body, delta function interaction describing the
scattering wave functions and the bound ($N$-body) states
\cite{McG64}.  This model has been generalized, examined by means of
several formalisms, or applied for different purposes
\cite{Yang68,Dodd,McG72,CD,McG88,Ger87}.  As stated before, the
attractive case for bosons has not been examined with periodic
boundary conditions and the present work fills this lacuna for $N=3$.

\section{Model description}

The stationary Schr\"odinger equation for three equal mass particles
in 1-dimension with 2-body delta function interactions reads
\begin{equation}
-\frac{\hbar^2}{2m}\sum_{i=1}^{3}
\frac{\partial^2\psi}{\partial y_i^2}
+2\tilde{c}\sum_{i<j} \delta(y_i-y_j)\psi=
\widetilde{E}\psi,
\end{equation}
where $y_i$ $(i=1,2,3)$ are the particle coordinates.
If they are enclosed in a box with length $L$ it is convenient   
to divide this equation by $\hbar^2/(2mL^2)$ and use instead
\begin{equation}\label{sch}
-\sum_{i=1}^{3}\frac{\partial^2\psi}{\partial{x}_i^2}
+2{c}\sum_{i<j} \delta({x}_i-{x}_j)\psi=
{E}\psi\,,
\end{equation}
expressed in terms of the dimensionless quantities
\begin{eqnarray}
x_i&=&y_i/L\\
E&=&2mL^2\widetilde{E}/\hbar^2\\
c&=&2m\tilde{c}L/\hbar^2.\label{c}
\end{eqnarray}
Note that $c$ is the only parameter of the model, and that, according
to (\ref{c}), the effect of enlargening the box or making the
interaction stronger are equivalent.  This work addresses the
attractive case in particular, corresponding to $c\le 0$.  But the
repulsive case ($c\ge0$) is also covered since the nature of the
solutions for the repulsive case are the same as one class of
solutions for the attractive case.  Of particular interest is how the
wavenumbers associated with a wavefunction change, as the potential
parameter $c$ varies continuously from repulsion to attraction.

        The delta function potential produces a jump in the
derivatives of the wave function where two particles meet.  This jump
is proportional to $c$ and to the wave function at that point,
\begin{equation}\label{dbc}
\left(\frac{\partial\psi}{\partial x_j}
-\frac{\partial\psi}{\partial x_k}\right)_{x_j=x_k+}
-\left(\frac{\partial\psi}{\partial x_j}
-\frac{\partial\psi}{\partial x_k}\right)_{x_j=x_k-}
=2c\,\psi|_{x_j=x_k}\,.
\end{equation}
Since the delta function interaction allows the particles to cross
each other, all orderings are possible, each ordering corresponding to
a particular ``region'' of coordinate space and one of the
permutations of the three particles.  On the basis that the particles
are bosons, it is sufficient to study the wave function in only one of
these regions, specifically the ``primary'' region
\begin{equation}
R_{123}: 0\le x_1\le x_2 \le x_3 \le 1.
\end{equation}
The wave function in any other region $R_{ijk}$ of coordinate space
is then simply obtained from the wave function in $R_{123}$ by
interchanging the particle labels.  In region $R_{123}$, the
equations (\ref{sch}) and (\ref{dbc}) can be written as
\begin{equation}\label{hel}
-\sum_{i=1}^3\frac{\partial^2}{\partial x_i^2}\psi=E\psi
\end{equation}
for $x_1\ne x_2\ne x_3\ne x_1$, and the jump conditions
\begin{equation}\label{bcd}
\left(\frac{\partial}{\partial x_{j+1}}-\frac{\partial}{\partial x_j}
\right)\psi\bigg|_{x_{j+1}=x_j}=c\psi\bigg|_{x_{j+1}=x_j}.
\end{equation}
If periodic boundary conditions are also imposed, a displacement of
$1$ in any of the coordinates $x_j$ leaves the function unchanged. In
particular, in $R_{123}$ this means that the wave function obeys
\begin{equation}\label{bcp1}
\psi(0,x_2,x_3)=\psi(x_2,x_3,1).
\end{equation}
Similarly, the derivatives satisfy
\begin{equation}\label{bcp2}
\frac{\partial}{\partial x}\psi(x,x_2,x_3)|_{x=0}=
\frac{\partial}{\partial x}\psi(x_2,x_3,x)|_{x=1}\,.
\end{equation}
In a two-body collision between particles of equal mass, the delta
function interaction can only interchange the momenta of the incident
particles or leave them unchanged. In other words, there is no
diffraction, that is, no ``new'' momenta, different from the initial
ones, are created.  For three equal mass particles with delta two-body
interactions, the eigenstates can thus be written in terms of only
three plane waves with (dimensionless) ``momenta'' $\{k_j\}$,
\begin{equation}\label{Bet}
\psi(x_1\le x_2\le x_3)=\sum_P a(P)
P \exp\left(i\sum_{j=1}^3 k_j x_j\right)\,,
\end{equation}
where the sum is over all permutations $P$ of the $\{k_j\}$, and
$a(P)$ are coefficients to be fixed by the boundary conditions
determined by the periodicity, Eqs. (\ref{bcp1}) and (\ref{bcp2}), and
by the delta function interaction, Eq. (\ref{bcd}). This wave function
form is known as the ``Bethe ansatz'' and it was first applied to spin
chains \cite{Be}. In the context of the Bose gas, Yang and Yang
\cite{Yang69} used a continuity argument to show that for positive $c$
{\em all} states are given by (\ref{Bet}) with real $k$'s, a result
which was later established rigorously by Dorlas \cite{Dorlas}. In the
present work it is shown that all states for $c\ge0$ are continuously
connected to $c\le0$ states in $k$ space so we are confident that the
eigenstates discussed later form in fact a complete set. An important
difference with the repulsive case is that for $c<0$ the $k$'s may
become complex.

        From (\ref{hel}) and (\ref{Bet}) the energy is simply obtained
as
\begin{equation}
E=\sum_{j=1}^{3}k_j^2\,,
\end{equation}
but this should not be interpreted as purely kinetic energy since
there is generally a potential energy contribution to $E$.  Note that
Eq. (\ref{hel}) is the Schr\"odinger equation only when the positions
of the particles are all different.  The true kinetic energy has to
take into account the jumps in the wave function derivative at the
region boundaries. The calculation of the potential energy is however
somewhat involved and is discussed in the Appendix.

        The structure of the coefficients $a(P)$ is imposed by the
jump boundary condition (\ref{bcd}) as explained, e.g., in \cite{LL}:
The amplitudes for two permutations differing by a transposition of
two particles are related by a factor $-e^{i\theta_{j\ell}}$,
\begin{eqnarray}
a(123)&=&1\nonumber\\
a(213)&=&-e^{i\theta_{21}}\nonumber\\
a(132)&=&-e^{i\theta_{32}}\nonumber\\
a(321)&=&-e^{i(\theta_{21}+\theta_{31}+\theta_{32})}\nonumber\\
a(312)&=&e^{i(\theta_{31}+\theta_{32})}\nonumber\\
a(231)&=&e^{i(\theta_{21}+\theta_{31})}\label{as}\\
\end{eqnarray}
where
\begin{equation}\label{etheta}
e^{i\theta_{j\ell}}\equiv\frac{c-i(k_j-k_\ell)}{c+i(k_j-k_\ell)}\,.
\end{equation}
By substituting (\ref{Bet}) and (\ref{as}) into the periodicity
conditions (\ref{bcp1}) and (\ref{bcp2}), the following set of coupled
transcendental equations is found
\begin{equation}\label{eik}
e^{-ik_j}=\exp\left[i\sum_{s=1}^3\theta_{sj}\right],\;\;\;\;j=1,2,3\,,
\end{equation}
where, by convention, $\theta_{jj}=0$.  Solving Eq. (\ref{eik}) for
the $k_j$ gives
\begin{eqnarray}\label{ktheta}
k_1&=&2\pi m_1-\theta_{21}-\theta_{31}\nonumber\\
k_2&=&2\pi m_2-\theta_{12}-\theta_{32}\nonumber\\
k_3&=&2\pi m_3-\theta_{23}-\theta_{13},
\end{eqnarray}
for some set of integers $\{m_j\}$, while the $\theta_{j\ell}$ are
given in terms of the $k_j$ by
\begin{equation}\label{thetak}
\theta_{j\ell}=i\ln\left[\frac{c+i(k_j-k_\ell)}{c-i(k_j-k_\ell)}
\right]=-2\arctan\left({k_j-k_\ell\over c}\right)\,.
\end{equation}
While it may appear to be natural to choose the principal branch of
the logarithm and the arctangent, another choice is more appropriate.
Since a state is uniquely defined by the set of numbers $\{k_j\}$,
irrespective of the order, because the particles are bosons, it is
convenient to order the $\{k_j\}$, when they are real, according to
\begin{equation}\label{order}
k_1\le k_2\le k_3\,.
\end{equation}
Consistent with this, the ranges of $\theta_{21},~\theta_{32}$ and
$\theta_{31}$ are chosen to satisfy
\begin{equation}
-2\pi<\Re(\theta_{j\ell})\le0,
\end{equation}
on the basis that the corresponding $k_j-k_\ell$ are positive.  In
this way, solving for the set of $\{k_j\}$ is equivalent to solving
for the set of $\theta_{j\ell}$ and it is noted that, for the above
choice for the range of $\theta_{j\ell}$, the $\theta_{j\ell}$ vary
continuously as $c$ and/or the $k_j$'s vary continuously over their
allowed ranges. This is also true if some of the $k_j$ and/or
$\theta_{j\ell}$ become complex, as discussed in latter sections.
Thus the $m_j$ provide a unique classification of the energy levels,
and for a given set of $m_j$, the energy eigenvalue (and eigenvector)
can be followed continuously as a function of $c$, as $c$ varies from
$\infty$ to $-\infty$.

        On taking the product of the three equations of the form of
Eq. (\ref{eik}), it follows that the total momentum $p$ is quantized,
\begin{equation}
p\equiv\sum_j k_j=2\pi(m_1+m_2+m_3)=2\pi n_p\,.
\end{equation}
This is the eigenvalue of the corresponding total momentum operator,
which commutes with the hamiltonian $H$.  It is clear that $p$ is an
invariant to the ``motion'' of an eigenvalue as $c$ varies
continuously from $\infty$ to $-\infty$.  For each solution set
$\{k_j\}$ there is another set $\{k_j'\}$ that also solves (\ref{eik})
and is related to the former by
\begin{equation}\label{strip}
k_j'=k_j+2\pi n_0\,.
\end{equation}
The transformation $\{k_j\}\to\{k_j'\}$ amounts to shifting the total
momentum by $6\pi n_0$.  This means that any state can be mapped to
another state in the central momentum strip $-3\pi<p\le3\pi$, and vice
versa, by such a transformation.  Thus we shall limit ourselves to
study only those states having total momentum in this strip, namely,
$n_p=-1,0,1$.

  A convenient set of variables, especially when the three $k$'s
are real, that uniquely define the state is $p$, $\delta_1$ and $\delta_2$,
where $\delta_1$ and
$\delta_2$ are, respectively, the relative momenta between particles
$12$ and $23$, namely
\begin{eqnarray}
\delta_1&\equiv&k_2-k_1\\
\delta_2&\equiv&k_3-k_2\,.
\end{eqnarray}
Note that, when real, the order assumed for the $k_j$,
Eq.(\ref{order}), implies that $\delta_j\ge0$. The $k$'s
are given in terms of these variables by
\begin{eqnarray}\label{k1}
k_1&=&\frac{1}{3}(p-2\delta_1-\delta_2) \\\label{k2}
k_2&=&\frac{1}{3}(p+\delta_1-\delta_2) \\\label{k3}
k_3&=&\frac{1}{3}(p+\delta_1+2\delta_2)\,.
\end{eqnarray}
and the energy takes the form
\begin{equation}\label{endel}
E=\frac{1}{3}[p^2+2(\delta_1^2+\delta_2^2+\delta_1\delta_2)]\,.
\end{equation}
Combinations of Eqs. (\ref{ktheta}) and (\ref{thetak}) give the
two coupled equations for $\delta_1$ and $\delta_2$,
\begin{eqnarray}\label{main1}
\delta_1&=&i\ln\left[\left(\frac{c+i\delta_1}{c-i\delta_1}\right)^2
\frac{c-i\delta_2}{c+i\delta_2}\,\,
\frac{c+i(\delta_1+\delta_2)}{c-i(\delta_1+\delta_2)}\right]
+2\pi n_1
\\
\label{main2}
\delta_2&=&i\ln\left[\left(\frac{c+i\delta_2}{c-i\delta_2}\right)^2
\frac{c-i\delta_1}{c+i\delta_1}\,\,
\frac{c+i(\delta_1+\delta_2)}{c-i(\delta_1+\delta_2)}\right]
+2\pi n_2\,.
\end{eqnarray}
where, on the basis that the principal part of the logarithm is taken,
$n_1$ is not necessarily equal to $m_2-m_1$, nor is $n_2$ necessarily
equal to $m_3-m_2$. In fact, unlike the $m_j$, the $n_j$ do not have
to remain constant as a given ``root'' of the coupled equations
$\{\delta_1(c),\delta_2(c)\}$ changes continuously with a variation of
$c$. That is why we shall not classify the roots according to
``local'' values $\{n_1(c),n_2(c)\}$, but according to their values
$\{n_1^0,n_2^0\}$ for no interaction, namely for $c=0$.  These are
unambiguously related to the set of quantum numbers $m_j$, see Eqs.
(\ref{n1}) and (\ref{n2}) below. If, as $c$ changes, the argument of
one of the logarithms, say $z_j$, crosses the negative real axis
(which is the branch cut for the principal part of the logarithm), its
phase changes abruptly by $\pm2\pi$ and the corresponding $n_j$ has to
jump up or down by one unit in order to follow the root continuously.
Of course, these discontinuities have no physical consequence and
merely reflect the choice made for the branch of the logarithm.  For
several formal manipulations and in particular for the object of
obtaining asymptotic expressions, it is useful to avoid the
discontinuities by continuing analytically the logarithm across the
branch cut, i.e., by passing to the contiguous Riemann sheet when the
argument $z_j$ crosses the real, negative axis. This will be discussed
further in Sec. IV to clarify the trajectories of the $k_j$ as $c$
changes smoothly.
 
         There are certain symmetries of the parameterization
$\delta_1, \delta_2$ and $p$ which can lead to energy degeneracies.
The interchange $\delta_1\rightleftharpoons\delta_2$ together with the
change in sign of $p$ are equivalent to the changes
$k_1\rightleftharpoons-k_3$ and $k_2\rightleftharpoons-k_2$ which, for
real $k$'s, inverts the order of Eq. (\ref{order}) to
$-k_1>-k_2>-k_3$.  This is also equivalent to taking the complex
conjugate of the wavefunction.  Thus, if $k_1\ne-k_3$ and/or
$k_2\ne0$, these are two different states with the same energy, a
two-fold degeneracy.  On the other hand, if $k_1=-k_3$ and $k_2=0$,
which is the special case that $\delta_1=\delta_2$ and $p=0$, then
this symmetry reproduces the same state, the wavefunction is real, and
the state is nondegenerate.  In terms of the classification of the
states of the central momentum strip by the $n_j^0$, it follows that
interchanging $n_1^0$ and $n_2^0$,  $(n_1^0\ne n_2^0)$, amounts to the
interchange of $\delta_1$ and $\delta_2$, and, see Eq. (\ref{totmo})
below, to change the sign of $p$, so that all signs of the $k$'s are
changed and the complex conjugate state is obtained. But for the
``diagonal'' case, $n_1^0=n_2^0$, there is no degeneracy. These states
are real and even under the parity transformation $x_j\to -x_j$.

        A second symmetry of the parameterization $\delta_1$,
$\delta_2$, $p$ is the interchange
$\delta_1\rightleftharpoons-\delta_2$ while $p$ remains unchanged.
This is equivalent to the changes $k_1\rightleftharpoons k_3$ and
$k_2\rightleftharpoons k_2$, which inverts the order of Eq.
(\ref{order}) to $k_1>k_2>k_3$ but does not change the momenta
themselves. But since the order of the $k$'s is immaterial, this is
just another way of labeling the same state.  In terms of the $n_j^0$
this means that $(n_1^0,n_2^0)$ and $(-n_2^0,-n_1^0)$ are actually the
same state.
 
        Because of the stated symmetry relations and the fact that any
eigenstate can be  translated by the total momentum shift
(\ref{strip}) to the central momentum strip, an exhaustive analysis of
all possible states is achieved by examining the cases $n_2^0\ge
n_1^0\ge 0$, since any other case is either equivalent to one of them
or obtained by a simple transformation. An understanding of the
behaviour of the roots and their limiting properties for different
ranges of $c$ requires a detailed analysis of how to carry out the
analytical continuation of the logarithms in Eqs.
(\ref{main1},\ref{main2}) as $c$ varies.  This is provided in the
following sections.  To keep track of the global picture a handy
summary of the results is provided in the final Section, and a set of
figures illustrate the essential aspects.

\section{The reference case of ``no interaction''}

   For $c=0$ there is no interaction and the particles move
freely.  In this case
\begin{equation}
e^{i\theta_{j\ell}}=-1\,,
\end{equation}
and with the present choice for the range of the $\theta_{j\ell}$,
\begin{equation}
\theta_{21}=\theta_{32}=\theta_{31}=-\pi.
\end{equation}
It follows that
\begin{eqnarray}
a(ijk)&=&1
\\\label{n1}
\delta_1=2\pi n_1^0&=&2\pi(m_2-m_1-1)
\\\label{n2}
\delta_2=2\pi n_2^0&=&2\pi(m_3-m_2-1)\,.
\end{eqnarray}
Equivalently, the $k_j$ are all real multiples of $2\pi$. Due to the
conventional order (\ref{order}), only the case where $n_j^0\ge 0$
needs to be considered in order to account for all the states of the
system.

        The total momentum in the central momentum strip for a state
$(n_1^0,n_2^0)$ is determined by noticing, from ({\ref{k2}), that if
$k_2$ is to be a multiple of $2\pi$, then $p+\delta_1-\delta_2$ has to
be a multiple of $6\pi$. For any pair $n_1^0$, $n_2^0$, and $p=2\pi
n_p\,\, (n_p=0,\pm 1)$, there is only one possible solution for $n_p$,
namely
\begin{equation}\label{totmo}
n_p=\cases{ 0 &if $\,\,n_1^0-n_2^0=3n$\cr
-1&if $\,\,n_1^0-n_2^0=3n+1$\cr
1&if $\,\,n_1^0-n^0_2=3n+2$\cr}\qquad n=0,\pm 1,\pm 2,\ldots\,.
\end{equation}

     For the states where one of the $n_j^0$ is zero, two $k$'s are
equal.  The equality of two $k$'s can occur only at $c=0$ and at
certain critical $c$ values discussed in Sec. V. (In general, for
$c\ne 0$, the wave function vanishes if two $k_j$ are equal.)  For the
ground state, $n_1^0=n_2^0=0$, the three $k_j$ are equal and the wave
function is a constant.

        The classification scheme used in the remainder of this paper,
for the energy eigenvalues and states, is based on following
$\delta_1$ and $\delta_2$ as continuous functions of $c$ from the
reference non-interacting system.  A symbol such as $(1,2)$ gives the
values of the quantum numbers $n_1^0$ and $n_2^0$ and identifies a
given ``root'' $\{\delta_1(c)$, $\delta_2 (c)\}$ of the transcendental
equations and the corresponding eigenstate (within the central strip
of total momentum) irrespective of the value of $c$.  Note that the
three ``quantum numbers'' $n_1^0,~n_2^0,~n_p$ are equivalent to the
set of quantum numbers $\{m_j\}$.  The total momentum given in
(\ref{totmo}) is independent of the value of $c$, as well, the
degeneracy of a root is also an invariant to the ``motion'' of the
root with $c$.

\section{Real $\lowercase{k}$ solutions}

        On the assumption that the $k_j$ all remain finite (and real),
it follows that
\begin{equation}\label{thetalims}
\left.\parbox{0.2in}{$\theta_{21}\\ \theta_{32}\\ \theta_{31}$}
\right\}\to\cases{0&if $c\to\infty$;\cr -2\pi&if $c\to-\infty$,\cr}
\end{equation}
with the consequence, see Eq. (\ref{ktheta}), that
\begin{eqnarray}\label{klimits}
        -\infty\leftarrow&c&\to\infty\nonumber\\
2\pi(m_1+2)\leftarrow&k_1&\to2\pi m_1\nonumber\\
2\pi m_2\leftarrow&k_2&\to2\pi m_2\nonumber\\
2\pi(m_3-2)\leftarrow&k_3&\to2\pi m_3.
\end{eqnarray}
Bounds on the $k$ differences can be narrowed by examining separately,
the detailed properties of the $\theta$'s for positive and negative
$c$, by the following reasoning:  On the basis of the chosen order for
the $k_j$, Eq. (\ref{order}), it follows that
\begin{equation}
k_3-k_1\ge k_2-k_1,\ k_3-k_2.
\end{equation}
As a consequence, for $c>0$ and $-\pi\le\theta_{j\ell}\le0$ for
$(j>\ell)$,
\begin{equation}
\tan\left({\theta_{31}\over2}\right)\le\tan\left({\theta_{21}\over2}
\right),\ \tan\left({\theta_{32}\over2}\right)
\end{equation}
and
\begin{equation}
\theta_{31}\le\theta_{21},\ \theta_{32},
\end{equation}
so that
\begin{equation}
-\pi\le\theta_{31}-\theta_{32}\le0
\end{equation}
and
\begin{eqnarray}\label{deltabounda}
2\pi n_1^0-\pi\le2\pi(m_2-m_1)-\pi+2\theta_{21}\le\delta_1&=&k_2-k_1=
2\pi(m_2-m_1)+2\theta_{21}+\theta_{31}-\theta_{32}\nonumber\\&&\le
2\pi(m_2-m_1)+2\theta_{21}\le2\pi(n_1^0+1).
\end{eqnarray}
An analogous argument for $c<0$ leads to
\begin{equation}\label{deltaboundb}
2\pi(n_1^0-1)<\delta_1=k_2-k_1\le2\pi n_1^0 +\pi,
\end{equation}
with the lower bound approached according to Eq. (\ref{klimits}).
Bounds for $\delta_2$ involving $n_2^0$ have the same structure.

        It is thus seen that if $n_1^0\ge2$ and $n_2^0\ge2$, then
these $k$ differences remain positive as $c$ varies from $\infty$ to
$-\infty$, with the consequence that the $k_j$ are real for all $c$
under these conditions.  This also implies that the energy tends to a
constant value for $|c|\to \infty$, see Fig. 1.  For these states, the
delta function potential acts, for both very large positive and
negative $c$, effectively as an impenetrable barrier.  That is, the
wave function at the region boundaries, $x_i=x_j$, tends to zero. This
can also be deduced from the jump conditions, Eq. (\ref{bcd}), since
the derivatives produce only the finite $k_j$ while $c$ becomes
infinite, so that consistency requires that $\psi$ vanishes.  The
potential energy as a function of $c$ decreases from zero to an
intermediate minimum and then grows again towards zero as
$c\to-\infty$ and vice versa for $c>0$, see the Appendix and Fig. 2.

        The above deduced limitations on how the $\delta$'s change
with $c$ allow a more detailed discussion of these changes from the
point of Eqs. (\ref{main1}) and (\ref{main2}).  An appropriate
starting point is to consider the limit of these equations when
$c\to0$, namely  
\begin{eqnarray}\label{main3}
\delta_1&=&i\ln\left[\left(\frac{\delta_1-ic}{\delta_1+ic}\right)^2
\frac{\delta_2+ic}{\delta_2-ic}\,\,
\frac{(\delta_1+\delta_2)-ic}{(\delta_1+\delta_2)+ic}\right]
+2\pi n_1^0
\\
\label{main4}
\delta_2&=&i\ln\left[\left(\frac{\delta_2-ic}{\delta_2+ic}\right)^2
\frac{\delta_1+ic}{\delta_1-ic}\,\,
\frac{(\delta_1+\delta_2)-ic}{(\delta_1+\delta_2)+ic}\right]
+2\pi n_2^0\,,
\end{eqnarray}
where the factors in the arguments have been rewritten to emphasize
that the $\delta$'s are in this limit the leading terms.  The
identification of the $n_j^0$ is on the basis that, on expanding these
equations as $c\to0$, Eqs. (\ref{n1}) and (\ref{n2}) are obtained,
namely
\begin{equation}\label{d0}
\delta_1=2\pi n_1^0+\frac{2n_1^0n_2^0+2(n_2^0)^2-(n_1^0)^2}
{n_1^0n_2^0(n_1^0+n_2^0)\pi}\,c+\cdots\,,
\end{equation}
with a symmetrical expression for $\delta_2$ on interchanging $n_1^0$
and $n_2^0$.  These expansions are valid only if both $n_j^0\ne0$,
while the exceptional cases are examined in Secs. VI and VII.

        As $c$ increases positively [always assuming the order of Eq.
(\ref{order})], the phases of the factors in the arguments of the
logarithms change, compare the $\theta_{j\ell}$ of Eq. (\ref{thetak})
and the discussion of their behaviour.  On tracing this behaviour as
$c\to+\infty$, it is seen that the phase of the argument of each
logarithm decreases by $2\pi$, so this can be taken into account when
changing Eqs. (\ref{main3}) and (\ref{main4}) into Eqs. (\ref{main1})
and (\ref{main2}) by setting $n_j(c\to \infty)=n_j^0+1$. For negative
$c$ the phase changes are the opposite and Eqs. (\ref{main3}) and
(\ref{main4}) are appropriately changed into
\begin{eqnarray}\label{main5}
\delta_1&=&i\ln\left[\left(\frac{-c-i\delta_1}{-c+i\delta_1}\right)^2
\frac{-c+i\delta_2}{-c-i\delta_2}\,\,
\frac{-c-i(\delta_1+\delta_2)}{-c+i(\delta_1+\delta_2)}\right]
+2\pi(n_1^0-1)
\\
\label{main6}
\delta_2&=&i\ln\left[\left(\frac{-c-i\delta_2}{-c+i\delta_2}\right)^2
\frac{-c+i\delta_1}{-c-i\delta_1}\,\,
\frac{-c-i(\delta_1+\delta_2)}{-c+i(\delta_1+\delta_2)}\right]
+2\pi(n_2^0-1)\,.
\end{eqnarray}
These are appropriate for expansions when $c\to-\infty$, but may of
course be used for all $c$ by analytic continuation.  In the same
vein, the related pair of equations with $n_j=n_j^0+1$ may be regarded
as valid for all $c$ by analytic continuation across the logarithm
branch cut.

        In summary, provided both $n_j^0\ge0$, then
\begin{equation}\label{cto+inf}
\delta_j{\buildrel c\to\infty \over\longrightarrow}2\pi(n_j^0+1)
\left(1-{6\over c}\right)+O(c^{-2}),
\end{equation}
while, provided both $n_j^0>1$, then
\begin{equation}
\delta_j{\buildrel c\to-\infty \over\longrightarrow}2\pi(n_j^0-1)
\left(1+{6\over-c}\right)+O(c^{-2}).
\end{equation}
In contrast, if $n_1^0=1$, then $\delta_1$ can approach 0 for finite
negative $c$, see Fig. 3.  For more negative values of $c$, the
$k_j$'s can become complex.  Similarly for $n_2^0=1$. Section V
discusses this situation.  If one of the $n_j^0=0$, then $\delta_j$
vanishes at $c=0$, see Sec. VI.

The case in which $n_1^0=n_2^0$ is particularly simple to analyze.  It
follows from Eqs. (\ref{main3}) and (\ref{main4}), analytically
continued for all $c$, that $\delta_1$ and $\delta_2$ satisfy the same
equation and thus are equal, with the consequence that the three $k$'s
for such an eigenstate remain equally spaced as $c$ varies.  After
dropping the subscripts, and formally written for $c>0$, the equation
for the common $\delta$ is
\begin{equation}\label{equaldel}
\delta=i\ln\left[\frac{(c+i\delta)(c+2i\delta)}
{(c-i\delta)(c-2i\delta)}\right]+2\pi(n^0+1).
\end{equation}
Fig. 4 illustrates how $\delta$ varies with $c$.  By implicit
differentiation,
\begin{equation}\label{dddc}
\frac{d\delta}{dc}=\frac{6\delta(c^2+2\delta^2)}
{c^2(c^2+5\delta^2)+4\delta^4+6c(2\delta^2+c^2)}\,.
\end{equation}
If $n_1^0=n_2^0\ne0$, Eq. (\ref{d0}) is consistent with the $c\to0$
limit of this result.

\section{States with at least one $\lowercase{n_j}^0=1$; Dimer states}

        If one $n^0_j=1$, for definiteness $n_1^0=1$, then from Eqs.
(\ref{deltabounda}) and (\ref{deltaboundb}), there is a possibility
that $\delta_1$ becomes $0$ for some critical negative value
$C(1,n_2^0)$ of $c$.  Note that the equations for $\delta_1$ and
$\delta_2$ are independent of $p$, so the critical value $C(1,n_2^0)$
has nothing to do with the value of $p$.  [The discussion for the
degenerate partner $(n_1^0,1)$ follows similar lines substituting
$n_1^0$ by $n_2^0$, $\delta_1$ by $\delta_2$ and $p$ by $-p$, see the
discussion of the symmetries of this parameterization in Sec. II.] At
a critical point $C(1,n_2^0)$, $k_1$ and $k_2$ become equal (if
$n_2^0=1$ as well, all three $k$'s become equal) and the Bethe ansatz
(\ref{Bet}) form for the wavefunction is no longer valid because it
vanishes.  Of course the normalization constant also vanishes so that
the normalized wavefunction does not vanish, but merely has a
different functional form obtained by taking the limit as $c\to
C(1,n^0_2)$ using the rule of l'Hospital.  A similar case was found by
Lieb and Liniger for one particular root in the case $N=2$ \cite{LL}.

        Two different cases arise, according to whether $n_2^0>1$, or
if $n_1^0=n_2^0=1$.  These cases are discussed in turn.

\subsection{The case when $\lowercase{n}_2^0>1$.}

        The value of the critical point $C(1,n_2^0)$ and the behaviour
in its neighborhood can be obtained by expanding Eqs. (\ref{main5})
and (\ref{main6}) for $\delta_1\to0$.  For such a purpose it is useful
to introduce the factors $u_j\equiv\delta_j/c$, $j=1,2$, and define
$u_0\equiv u_2(c=C)$ as the critical value of $u_2$.  After fairly
extensive algebra, it follows that
\begin{equation}
c=C(1,n_2^0)+{21+24u_0^2+8u_0^4\over6(1+u_0^2)^2}u_1^2+\cdots
\end{equation}
with
\begin{equation}\label{ccrit}
C(1,n_2^0)=-4-{2\over1+u_0^2},
\end{equation}
where $u_0$ is determined by the
transcendental equation
\begin{equation}
3i\ln{1+iu_0\over1-iu_0}+2\pi(n_2^0-1)+4u_0+{2u_0\over1+u_0^2}=0.
\end{equation}
In the neighborhood of the critical point, $u_2$ changes according to
\begin{equation}\label{u2u1}
u_2=u_0-{1\over2}u_1+{3+6u_0^2+2u_0^4\over6u_0(1+u_0^2)}u_1^2
+\cdots\,.
\end{equation}
It is seen that $\delta_1(c)$ has a square root singularity at the
critical point, being real for $c>C(1,n_2^0)$ and pure imaginary for
$c<C(1,n_2^0)$.  An analytic connection between real and imaginary
branches of $\delta_1$ can be made by attributing c with a small
imaginary part.  The association that is used in the following
parameterization is consistent with $c$ having a small negative
imaginary part, essentially adding a small dissipative contribution to
the Hamiltonian. It is also noticed from Eq. (\ref{ccrit}) that all
the critical values of $c$ lie between $-6$ and $-4$.  Actually, to
three decimal places, the lowest critical value is $C(1,2)=-4.163$,
increasing towards $-4$ as $u_0$, and $n_2^0$, increases. These
aspects are illustrated in Fig. 3. 

     For $c<C(1,n_2^0)$, $k_1$ and $k_2$ become a complex conjugate
pair, while $k_3$ remains real.  The discussion of the root in this
region is better examined by using a new set of real variables,
$\alpha$ and $\gamma$, defined by
\begin{eqnarray}\label{dag}
\delta_1&=&-2i\alpha \nonumber\\
\delta_2&=&i\alpha-3\gamma\,,
\end{eqnarray}
so that 
\begin{eqnarray}\label{kag}
k_1&=&i\alpha+\gamma+p/3 \nonumber\\
k_2&=&-i\alpha+\gamma+p/3 \nonumber\\
k_3&=&-2\gamma+p/3.
\end{eqnarray}
This is consistent with Eq. (\ref{u2u1}) for $\alpha\to0$, on the basis
that both $\alpha$ and $\gamma$ are real.  This parametrization could
of course be used for all $c$ [with $\alpha$ possibly imaginary],
since $\delta_1/2=-i\alpha$ has the physical meaning of the relative
momentum between particles 1 and 2 while $2\gamma$ is the relative
momentum between the pair $12$ and particle $3$,
\begin{eqnarray}
-i\alpha&=&\delta_1/2=\frac{k_2-k_1}{2}\\
2\gamma&=&k_3-{p\over3}=\frac{2}{3}\left(k_3-\frac{k_1+k_2}{2}\right).
\end{eqnarray}
For real $\alpha$, $1/\alpha$ gives a measure of the size of the
``bound states'' formed.  Of course, in a finite box all states are,
strictly speaking, bound, i.e., their energies are discrete and their
spatial extention is limited by the box length.  But when $\alpha$ is
real, the state is localized even further so that the probability of a
pair of particles being close together has been significantly
enhanced.  As well, the energy has a negative contribution, see
(\ref{enag}) below, so that an energy gap arises in the spectrum
between states with real or imaginary $\alpha$, see Fig. 1.  Since
these are all basic ingredients of proper bound states, this
terminology seems justified.

        With this parametrization, the total energy is decomposed into
separate quadratic contributions from the three variables,
\begin{equation}\label{enag}
E=-2\alpha^2+6\gamma^2+p^2/3\,,
\end{equation}
and the system of transcendental equations takes the form
\begin{eqnarray}
-2\alpha&=&\ln\left[\left(\frac{-c-2\alpha}{-c+2\alpha}\right)^2
\frac{-c-\alpha-3i\gamma}{-c+\alpha+3i\gamma}\,
\frac{-c-\alpha+3i\gamma}{-c+\alpha-3i\gamma}\right]
\label{ag1}\\
i\alpha-3\gamma&=&i\ln\left[\left(\frac{-c+\alpha+3i\gamma}
{-c-\alpha-3i\gamma}\right)^2
\frac{-c+2\alpha}{-c-2\alpha}\,
\frac{-c-\alpha+3i\gamma}{-c+\alpha-3i\gamma}\right]+2\pi(n_2^0-1)
\label{ag2}
\end{eqnarray}
which corresponds to the correct phase form for $c<0$, according to
Eqs. (\ref{main5}) and (\ref{main6}).  It is consistent to solve these
equations maintaining $\alpha$ and $\gamma$ real, which is also
consistent with the local behaviour, Eq. (\ref{u2u1}), as
$\alpha\to0$. Before entering into the detailed analysis of the
equations, it is worth examining Figures 5 and 6 to quickly visualize
the behaviour of these two parameters with $c$.  Solid lines
correspond to $n^0_1=0$ and dashed lines to $n^0_1=1$, whereas the
numbers close to the different lines give $n_2^0$.  As $c$ becomes
more negative $\alpha$ increases.  This concentrates the wavefunction
to where the particles are close together and $\langle V\rangle $ and
$\langle E\rangle$ become very large and negative, see in Figs. 1 and
2 the lines for $n=0,1$.  A prominent feature in Fig. 5 is the
grouping into two asymptotic behaviours for $\alpha$ as $c\to
-\infty$.  These will be later associated with dimer and trimer
configurations.  Note also that all states with $n_1^0=0$ have a
common critical point at $c=0$ (where $\alpha=0$), while for $n_1^0=1$
the critical points spread from $c=-6$ to $c=-4$. Another interesting
point is the quasi-invariance of $\alpha$ with respect to $n_2^0$ for
$n_2^0\ge 2$ at fixed $c$.  These states have essentially the same
binding strength and differ only by pair-single relative momentum, and
possibly by total momentum. Fig. 6 for $\gamma$ has a simpler
structure based on a rather regular pattern of lines with equally
spaced asymptotic values. Except for the two cases (0,0) and (1,1)
with $\gamma=0$ for all $c$, which corresponds to no relative motion
between a particle pair and a single particle, when $c$ becomes more
negative, $\gamma$ varies smoothly and tends to a constant value. When
$\gamma$ is essentially constant, $c$ changes the strength of the
attraction between the pair (i.e., the value of $\alpha$), but not the
motion of the third particle with respect to the pair.

        The structure of the spectrum of energy levels can be
described in terms of how the energy varies as the three quantum
numbers $n_1^0, n_2^0, n_p$ change.  This can be attributed to several
types of ``elementary excitations'' which are associated with
different physical effects. Within the central momentum strip, for
states with complex $k_j$'s and for $-c$ large, these are:\hfil\break
a) As $n_2^0$ varies, a change of relative pair-single motion by
$\Delta_\gamma\approx\pi/3$, with $\alpha$ essentially constant and
$|p|$ constant.  In Fig. 6 this is not possible between all contiguous
levels of $\gamma$, but only for those where $n_p=1$ and $n_{p'}=-1$,
see Eq. (37).  If one of the states has momentum zero, a jump to the
nearest level necessarily implies in addition to $\Delta_\gamma$, an
elementary total momentum jump $\Delta_p=2\pi$;
b) Transitions between a trimer and a dimer state or from a dimer to a
pair-absent state, with $\delta_\alpha\approx -c/2$;
c) A minimum total momentum jump by $2\pi$, with $\alpha$ and $\gamma$
constant.  This may only occur between the states (0,0) and (0,1) or
(1,0).  Any other transition changes $\gamma$.  But $\alpha$ and
$\gamma$ may also stay constant if the system changes to a different
momentum strip by a total momentum translation of $\Delta_p=6\pi$. Of
course multiples or combinations of these elementary excitations are
possible and complicate the spectrum considerably.

        The detailed quantitative features of $\alpha$ and $\gamma$ as
functions of $c$ are now examined.  On the basis that $\alpha$ and
$\gamma$ are real, the real part of Eq. (\ref{ag2}) is
\begin{equation}\label{ga}
\frac{3}{2}i\ln\left[\frac{(-c+\alpha-3i\gamma)(-c-\alpha-3i\gamma)}
{(-c+\alpha+3i\gamma)(-c-\alpha+3i\gamma)}\right]=3\gamma
+2\pi(n_2^0-1).
\end{equation}
This provides upper and lower bounds for $\gamma$, namely
\begin{equation}\label{gbound}
-(4n_2^0-1)\pi/6<\gamma<-(4n_2^0-7)\pi/6\,.
\end{equation}
Equation (\ref{ga}) may also
be written as
\begin{equation}\label{arct}
\arctan\left(\frac{6\gamma c}{c^2-9\gamma^2-\alpha^2}\right)
=-\gamma-2\pi(n_2^0-1)/3
\end{equation}
which is useful for the determination of the asymptotic behaviour of
$\alpha$ and $\gamma$.

      It is also possible to solve for $\gamma^2$ in Eq. (\ref{ag1})
in terms of $\alpha$ and $c$,
\begin{equation}\label{g2}
\gamma^2=\frac{e^{-\alpha}(c-2\alpha)^2(c-\alpha)^2
-e^{\alpha}(c+2\alpha)^2(c+\alpha)^2}
{9[e^{\alpha}(c+2\alpha)^2-e^{-\alpha}(c-2\alpha)^2]}\,.
\end{equation}
As $\alpha\to 0$, $\gamma^2(\alpha=0)=c^2(6+c)/[-9(4+c)]$.  This is
identical to Eq. (\ref{ccrit}) for the relation between $\delta_2$
and $c$ at a critical value of $c$.

        The asymptotic behaviour as $c\to-\infty$ is now
investigated.  According to Eq. (\ref{gbound}), $\gamma$ remains
finite while $\alpha$ satisfies Eq. (\ref{ag1}).  Both positive and
negative $\alpha$ are solutions to this equation, but since these just
correspond to an interchange of $k_1$ and $k_2$, only the positive
root is examined.  Eq. (\ref{ag1}) can be rewritten as an equation
involving only real quantities, namely
\begin{equation}\label{71}
-2\alpha=\ln\left[\left(\frac{-c-2\alpha}{-c+2\alpha}\right)^2
\frac{(-c-\alpha)^2+9\gamma^2}{(-c+\alpha)^2+9\gamma^2}\right].
\end{equation}
The obvious (but invalid) approach to try when making an asymptotic
expansion, is to expand in powers of $\gamma^2$ since the other
factors involve $c$. This implies that $\alpha$ must also remain
finite and leads to the requirement that $c\to-6$, an inconsistency.
It follows that either $(-c-2\alpha)\to0$ or $(-c-\alpha)\to0$ in the
Limit $c\to-\infty$. But it is noticed that as $c$ changes from its
critical value $C(1,n_2^0)$ to $-\infty$, $\alpha$ changes from $0$ to
its asymptotic behaviour, yet the factors in the argument of the
logarithm, $-c\pm 2\alpha$ and $(-c\pm\alpha)^2+9\gamma^2$, must
remain positive or an imaginary phase factor must be added to the
right hand side.  Since such a case would imply that $\alpha$ becomes
complex, this is not allowed. Another way to understand the
preservation of sign of all factors is that if one of them became zero
for a finite $c$ (and $\alpha$) the logarithm, and the left hand side
of the equation, would be infinite in absolute value, which is again
inconsistent with the finite value of $\alpha$ on the right hand side.
The only form of $\alpha$ that maintains all factors positive, is
$\alpha=-{1\over2}c+\beta$, with $\beta<0$ and $\beta/c\to0$
asymptotically.  A straightforward expansion then gives
\begin{eqnarray}
\beta&=&3ce^{c/2}-9c^2e^c+\cdots,\nonumber\\
\alpha&=&-{1\over2}c+3ce^{c/2}-9c^2e^c+\cdots,
\end{eqnarray}
and from Eq. (\ref{arct}),
\begin{equation}
\gamma=-{2\pi\over3}(n_2^0-1)\left[1-{8\over c}+\cdots\right],
\end{equation}
as the asymptotic expansions for both $\alpha$ and $\gamma$.

\subsection{The case when $\lowercase{n}_2^0=\lowercase{n}_1^0=1$.}

        As previously mentioned, this implies that $\delta_1=\delta_2$
for all $c$, and the common $\delta$ is determined by Eq.
(\ref{equaldel}), which is appropriate for $c>0$.  Reexpressing this
for $c<0$, retaining a continuous relation for the phase, gives
\begin{equation}
\delta=i\ln\left[\frac{(-c-i\delta)(-c-2i\delta)}
{(-c+i\delta)(-c+2i\delta)}\right],
\end{equation}
and taking into account that $n^0=1$.  Clearly $\delta=0$ is a
solution of this equation.  But to find the corresponding critical
value $C(1,1)$ of $c$ and the behaviour in the neighborhood of this
critical point, an expansion is needed.  This is easily accomplished
to yield
\begin{equation}
c=-6+{1\over6}\delta^2+\cdots\,.
\end{equation}
This immediately shows that the critical value is $C(1,1)=-6$, and
that there is a square root singularity of $\delta$ as a function of
$c$.  For $c<-6$, $\delta=-i\alpha$ is pure imaginary [negative
imaginary if the same connection around the singularity is used as in
the last subsection].  It follows that the $k_j$ are given in terms of
this parameterization by
\begin{eqnarray}\label{g0}
k_3&=&-i\alpha+p/3\nonumber\\
k_2&=&p/3\nonumber\\
k_1&=&i\alpha+p/3,
\end{eqnarray}
while $\alpha$ is determined by
\begin{equation}\label{alp}
\alpha=\ln\left(\frac{1-3\alpha'+2\alpha'^2}{1+3\alpha'
+2\alpha'^2}\right),
\end{equation}
where $\alpha'=\alpha/c$.  [Contrast this with Eq. (\ref{kag}).  The
associations made for the $k_j$ in (\ref{g0}) are more natural here,
since $\delta_1=\delta_2$ and the three $k_j$ change continuously
across the critical point, but in fact (\ref{kag}), with $\gamma=0$
could be used as well because the state is defined by the three
``momenta'' regardless of the ordering convention.]
  
        Clearly, if $\alpha'\to0$ as $c\to-\infty$, then $\alpha\to0$.
But on expanding to look at the correction terms, this assumption
implies that $c\to6$, an inconsistent result.  It follows that
$\alpha$ grows to $\infty$ asymptotically.  On the basis that only
$\alpha>0$ solutions are needed ($\alpha'<0$), the vanishing of the
denominator in the argument of the logarithm requires either
$\alpha'\to-1$ or $\alpha'\to-{1\over2}$ as $c\to-\infty$. Only the
latter is consistent with no accumulation of phase as $c$ changes from
$-6$ to $-\infty$, and the reality of $\alpha$. (Note that, by a
similar argument to the one below Eq. (\ref{71}), numerator and
denominator in (\ref{alp}) must preserve their sign as $c$ varies from
$-6$ to $-\infty$ which implies the bound $2\alpha<-c$.) On setting
$\alpha=-{1\over2}c+\eta$, it follows on expansion that
\begin{equation}
\alpha=-{1\over2}c+3ce^{c/2}+\cdots
\end{equation}
as the asymptotic behaviour of $\alpha$ in this case.

\section{States with one $\lowercase{n_j}^0=0$: Dimer and Trimer
States}

        This group of states are of the form $(0,n_2^0)$, and of
course their degenerate partners.  Only the case $(0,n_2^0)$ is
explicitly treated here since their corresponding partner states are
simply obtained by symmetry, specifically for the states when $c<0$ by
changing the sign of $\gamma$.  For $c\to\infty$, these states fit
into the formulation of Eq. (\ref{cto+inf}), so no further discussion
of this limit is needed.

         As $c\to0$, the starting point for the analysis is the pair of
equations, (\ref{main3}) and (\ref{main4}).  Since $n_1^0=0$,
$\delta_1$ and $c$ are to simultaneously approach $0$, and it is to be
expected by analogy with the other cases, that $c$ will be
proportional to $\delta_1^2$.  This is confirmed by the following
argument:  Since $\delta_1\to0$, then the argument of the logarithm in
Eq. (\ref{main3}) must approach 1.  For the ratios of the terms
involving $\delta_2$, these approach 1 since they are dominated by the
common non-zero value of $\delta_2$.  That leaves the square term. For
this to approach 1, each factor must be dominated by the common
$\delta_1$, which implies that it is the ratio $c/\delta_1$ that is
small and can be used as a variable for which the logarithm is
expanded, as well as the ratios $c/\delta_2$ and $\delta_1/\delta_2$,
all of which vanish as $c\to0$.  Explicitly, the expansions of Eqs.
(\ref{main3}) and (\ref{main4}) are
\begin{eqnarray}
\delta_1&=&4{c\over\delta_1}-{4c^3\over3\delta_1^3}
-2{c\delta_1\over\delta_2^2}+O(\delta_1^4)\nonumber\\
\delta_2&=&2\pi n_2^0-2{c\over\delta_1}+{2c^3\over3\delta_1^3}
+6{c\over\delta_2}-2{c\delta_1\over\delta_2^2}+O(\delta_1^4).
\end{eqnarray}
It follows that $c$ is proportional to $\delta_1^2$, as was to be
deduced.  Rearrangement of these series for $c\to0$ gives the
expansion in powers of $\delta_1$ as
\begin{eqnarray}
c&=&{1\over4}\delta_1^2+\left[{1\over192}+{1\over32(\pi n_2^0)^2}
\right]\delta_1^4+\cdots \nonumber\\
\delta_2&=&2\pi n_2^0-{1\over2}\delta_1+{3\delta_1^2\over4\pi n_2^0}
+\cdots\,.
\end{eqnarray}
For $c<0$, the parameterization of Eq. (\ref{dag}) is appropriate.
In the limit $c\to-0$, the ratio $c/\delta_1$ transforms according to
\begin{equation}
{c\over\delta_1}\to {1\over4}\delta_1\to-{1\over2}i\alpha\leftarrow
i{c\over2\alpha},
\end{equation}
with the consequence that Eqs. (\ref{main3}) and (\ref{main4}) become
\begin{eqnarray}\label{main7}
-2\alpha&=&\ln\left[\left({2\alpha+c\over2\alpha-c}\right)^2
{(\alpha+c)^2+9\gamma^2\over(\alpha-c)^2+9\gamma^2}\right]
\\
\label{main8}
i\alpha-3\gamma
&=&i\ln\left[\left({2\alpha-c\over2\alpha+c}\right)
\left({i\alpha-ic-3\gamma\over i\alpha+ic-3\gamma}\right)^2
\left({-i\alpha-ic-3\gamma\over-i\alpha+ic-3\gamma}\right)\right]
+2\pi n_2^0.
\end{eqnarray}
It is also important to note that, as a consequence of the analytic
continuation into the $c<0$ region, $2\alpha+c>0$, and
$\gamma\to-\delta_2/3<0$.  To avoid further singularities in the
logarithmic expressions (which would lead to inconsistency between
right and left sides of the transcendental equations), these
constraints must hold for all $c<0$.

        For exploring the behaviour as $c\to-\infty$, Eq.
(\ref{main7}) for $\alpha$ has an appropriate form for expansion about
the large $-c$ and $\alpha$.  But if one term in each of the factors
is to dominate the expansion, then the result would imply that the
logarithmic expression becomes finite, a result inconsistent with the
left hand side approaching $-\infty$.  As a consequence, since
$\alpha>-c/2$, either $\alpha\to-c/2$ or the combination,
$\alpha\to-c$ and $\gamma\to0$, must occur.  The first alternative is
now shown to be valid only for $n_2^0>1$, while the second alternative
is valid only if $n_2^0=1$.

        The limiting case $\alpha=-{1\over2}c+\beta$ leads to a
straightforward expansion of Eq. (\ref{main7}),
\begin{equation}
-2\alpha=c-2\beta=2\ln\left({\beta\over-3c}\right)+{22\beta\over3c}
+\cdots\,,
\end{equation}
which can be rewritten as the equation
\begin{equation}
\beta=-3ce^{c/2}+9c^2e^c+\cdots
\end{equation}
for $\beta$.  As $c\to-\infty$, this vanishes exponentially.  The
behaviour of $\gamma$ is to be obtained from Eq. (\ref{main8}).  But
the dominant quantity $c$ is multiplied by the phase factors $\pm i$,
so that an asymptotic expansion carries along a phase change for the
logarithm.  On carefully analyzing how the various factors change as
$c$ changes from 0 to $-\infty$, a phase change of $e^{3\pi i}$ in
the argument of the logarithm is found.  The real part of Eq.
(\ref{main8}) determines $\gamma$, which after expanding and rewriting
gives
\begin{equation}
\gamma=-\left({2\over3}n_2^0-1\right)\pi\left(1
-{8\over c}\cdots\right).
\end{equation}
Since it is required that $\gamma<0$, it is seen that this expansion
is only valid for $n_2^0>1$.

        The limiting case $\alpha=-c+\eta$ also allows a
straightforward expansion of Eq. (\ref{main7}), but in this case the
resulting equation for $\eta$ involves $\gamma$ in the lowest order
term, namely
\begin{equation}
\eta^2+\gamma^2=36e^{-2\eta}c^2e^{2c}\left(1-{5\eta\over3c}\cdots
\right).
\end{equation}
Since both $\eta$ and $\gamma$ are real, this implies that both these
quantities must vanish asymptotically as $c\to-\infty$.  In this case,
Eq. (\ref{main8}) accumulates a phase $e^{3\pi i/2}$ in the argument
of the logarithm when transforming the factors so that they will be
dominated by a positive real part in the limit. What is crucially
different in this case from the previous one is the presence of finite
complex factors $-3\gamma\mp i\eta$.  As a result, the real part of
Eq. (\ref{main8}) has the asymptotic expansion
\begin{equation}
-3\gamma={3\over2}\arctan\left({-\eta\over-3\gamma}\right)
+\left(2n_2^0-{3\over2}\right)\pi-{9\gamma\over2c}+\cdots\,.
\end{equation}
As $c\to-\infty$, $\gamma$ must vanish for this case, as was deduced
from the expansion of the $\alpha$ equation.  Thus in the limit, the
identity
\begin{equation}
\lim_{c\to-\infty}\arctan\left({\eta\over-3\gamma}\right)=
\left({4\over3}n_2^0-1\right)\pi
\end{equation}
must be satisfied.  Since the magnitude of the arctangent is bounded
by $\pi/2$, this identity can only be satisfied if $n_2^0=1$, in which
case the limiting ratio of $\eta$ and $\gamma$ is determined by the
condition $\eta=-3\gamma\tan(\pi/3)$.  In summary, for $n_2^0>1$ the
asymptotic behaviour of $\alpha$ is $\alpha\to-c/2$, which gives states
of dimer type, while for $n_2^0=1$, $\alpha\to-c$ and all three
particles are forced to be close to one another, a trimer state.
These associations will be discussed in Section VIII.  

\section{The Ground State: $\lowercase{n_1^0=n_2^0}=0$}

        Since $n_1^0$ and $n_2^0$ are equal, it follows that
$\delta_1=\delta_2\equiv\delta$.  Near $c=0$ and for $c>0$, the
appropriate equation for determining $\delta$ is Eq. (\ref{main3}),
which is modified for equal $\delta$'s to be
\begin{equation}\label{mainequal}
\delta=i\ln\left[\frac{(\delta-ic)(2\delta-ic)}
{(\delta+ic)(2\delta+ic)}\right].
\end{equation}
The behaviour as $c\to\infty$ is covered by the expansion of Eq.
(\ref{equaldel}), with asymptotic form identical to Eq.
(\ref{cto+inf}).  Since $\delta\to0$ as $c\to0$, it is necessary that
the argument of the logarithm must approach 1, which requires that
$c\to0$ faster than does $\delta$.  Thus the expansion parameter is
$c/\delta$, so that after expansion and rearrangement
\begin{equation}
c={1\over3}\delta^2+{1\over108}\delta^4+\cdots\,.
\end{equation}
For $c<0$, the parameterization of Eq. (\ref{g0}) is used, this being
equivalent to $\delta=-i\alpha$ with $\alpha>0$.  The ratio
$-ic/\delta$ for $c>0$ thus becomes $c/\alpha$ for $c<0$ and Eq.
(\ref{mainequal}) becomes
\begin{equation}\label{acln}
\alpha=\ln\left[\frac{(\alpha-c)(2\alpha-c)}
{(\alpha+c)(2\alpha+c)}\right].
\end{equation}
This is identical to Eq. (\ref{alp}), but now the  constraint is that
$\alpha>-c$.  This constraint requires that as $c\to-\infty$, $\alpha$
must become infinite and the expansion of the logarithm is about a
singular point of the logarithm.  The only possible form is
$\alpha=-c+\eta$, with $\eta>0$ approaching zero. After rearrangement,
the resulting expansion gives
\begin{equation}
\eta=-6ce^c-36c^2e^{2c}+\cdots\,.
\end{equation}
which is consistent with $\eta$ being positive.

\section{Types of states and their representation}

The states are best represented as contour plots of the  probability
density (and of the phase if required) in a ``ternary phase diagram''
for the variables
\begin{eqnarray}
r_{12}&=&x_2-x_1\nonumber\\
r_{23}&=&x_3-x_2\nonumber\\
r_{31}&=&1+x_1-x_3\,,
\end{eqnarray}
constrained to the region $R_{123}$, $0\le r_{ij}\le 1$.  Note that
these three coordinates always add to one.  In this diagram the
coordinate points are represented in an equilateral triangle.  Each of
the base lines corresponds to one of the coordinates $r_{ij}$ being
zero, and each point in the base line corresponds to a particular
location of the third particle at the right or left side of the pair
$ij$ (the closer to the vertex, the closer the third particle is to
the pair).  The lines parallel to the base are lines of constant
$r_{ij}$.  The value of $r_{ij}$ increases from zero at the base to
one at the opposite vertex (labeled as $r_{ij}$ in the figures).  The
center of the triangle is the point where the three distances are
equal to $1/3$.  Near the vertex $r_{ij}$ the distance between
particles $j$ and $i$ is also small (and tends to zero at the vertex
itself); the difference with the basis region is that now the third
particle is {\em between} the particles $j$ and $i$.
   
        In summary, bases are associated with two particles being
together (dimer configurations), and vertices with the three particles
being together (trimers).  However, in a general state with three real
$k$'s, there is no bias towards these configurations.  Recall that the
wave functions in $R_{123}$ are linear combinations of six
exponentials that can be obtained from the (123) form,
\begin{equation}\label{pw}
e^{i(k_1 x_1+k_2 x_2+k_3 x_3)}\,,
\end{equation}
by permuting the $k$'s in all possible manners.   The probability
density (square modulus) of any of these real $k$ plane wave terms is
constant; in other words, in these plane waves none of the particle
configurations is favored. The interference between the six different
plane waves however destroys the spatial homogeneity (except for the
ground state at $c=0$) and provides some structure with maxima and
minima, see Fig. 7.  The complex $k$ case is different, see Figs. 8
and 9 for examples of trimer and dimer states. Eq. (\ref{pw}) for the
$(123)$ exponential may now be written, using the parameterization
in (\ref{kag}), as
\begin{equation}
e^{i(x_1+x_2+x_3)p/3}e^{i\gamma(x_1+x_2-2x_3)}e^{\alpha(x_2-x_1)}\,,
\end{equation}
where the plane wave for the center of mass motion,  a plane wave for
relative motion of the pair 12 with respect to particle 3, and a real
exponential can be recognized; for $\alpha>0$ the exponent is positive
and it favors the trimer configuration of the vertex ($r_{12}=1$). By
permuting $k_1$ and $k_2$, one finds instead, in the term $(213)$, a
negative exponent that favors the dimer configurations $r_{12}=0$. Of
course the other two pairs, 13 and 23, have also a corresponding set
of dimer and trimer contributions, so that the six terms of the wave
function can be separated into two groups: Three of them, $(123)$,
$(231)$ and $(312)$, represent trimer configurations, and the other
three, $(213)$, $(132)$ and $(321)$, represent dimer configurations.
The relative weights among them are determined by the amplitudes
$a_{ijk}$.  For ``trimer states'' the three trimer terms dominate the
linear combination and the energy becomes, as $c\to-\infty$, the
energy of an actual trimer state (for three particles on an infinite
line). For ``dimer states'', there is also significant density along
the edges of the triangle (not only at the vertices), and the energy
tends to the energy of the actual dimer (on the infinite line), plus
the contributions from relative motion of the dimer with the free
particle and of the center of mass motion.  This is consistent with
our expectation of reproducing infinite line results in the limit of a
large box. As a concrete example, the (unnormalized) states (0,0) and
(1,1) for $c\le -6$ are examined: In both cases $\gamma=p=0$ and they
can be written, using Eqs. (\ref{Bet}), (\ref{as}), (\ref{etheta}) and
(\ref{acln}), as
\begin{eqnarray}
\psi&=&e^{\alpha r_{12}}+e^{\alpha r_{23}}+e^{\alpha r_{31}}
\nonumber\\ &&
+\frac{2\alpha-c}{2\alpha+c}\left(e^{-\alpha r_{12}}+
e^{-\alpha r_{23}}+e^{-\alpha r_{31}}\right).
\end{eqnarray}
As $c\to-\infty$ the factor multiplying the dimer terms in parenthesis
tends to $3$ for $(0,0)$ (which makes this contribution negligible)
but to $\infty$ for $(1,1)$. Fig. 8 shows the ground state $(0,0)$ for
negative $c$.

        An important aspect of these associations is that the trimer
or dimer character changes continuously along a given root as $c$
varies, and only asymptotically ($c\to-\infty$) is the separation
between trimer and dimer states unambiguous. For any finite negative
$c$ the complex $k$ roots have non-zero dimer and trimer components.
Note for example how the state $(0,1)$ goes from a dimer dominated
behaviour to trimer behaviour as $c$ becomes more negative in Fig. 5.
In the same vein, even though the threshold between real and complex
$k$ is well defined and it occurs at a critical value of $c$, there
aren't any dramatic (discontinuous) changes in the wave function, and
the energy varies smoothly with $c$, see Fig. 1, in the neighborhood
of the critical $c$ values. However, a different qualitative behaviour
(of the energy and state probability density) becomes clear when
comparing the state below and above the critical point as the distance
from $C$ increases.  Thus the critical values indicate a transition of
the root from one character, without pair formation, to another where
dimers or trimers can be recognized.

\section{Summary}

        A model of three bosons subject to delta function interactions
and periodic boundary conditions has been analyzed.  In particular a
description of the eigenstates and their behaviour has been given in
terms of three momenta $k_i$, $i=1,2,3$ or two sets of alternative
parameterizations, $\delta_1,\delta_2,p$ and $\alpha, \gamma, p$,
convenient, respectively, for the cases where the $k$'s are real or
complex. The roots can be primarily classified according to whether
the three momenta remain real, or not, for all $c$.  In the second
case the wave function tends to concentrate asymptotically around
dimer or trimer configurations and the energy decreases quadratically
with $c$ as $c\to-\infty$. The critical values of $c$ required to form
the bounds (go from real to complex $k$'s) have been provided.

        The main features of the root behaviour as $c$ varies are now
summarized. (Using the symmetry properties or total momentum
translations the behaviour of any other state is obtained from the
ones we consider explicitly, namely states $(n_1^0,n_2^0)$, $n_1^0\le
n_2^0$, in the central momentum strip, $n_p=0,\pm 1$.) Relations
satisfied by all roots are:
\begin{itemize}
\item The total momentum $p$ (within the central strip) is given from
$n_1^0$ and $n_2^0$ by (\ref{totmo}).  It is constant as $c$ varies
for a given root. The total energy varies according to (\ref{endel})
or (\ref{enag}).
\item $\delta_j(c=0)=2\pi n_j^0$. \item $\delta_j\to 2\pi(n_j^0+1)$ as
$c\to\infty$.
\end{itemize}
The different particular cases are
characterized by the following properties:
\begin{itemize}
\item (0,0): $C(0,0)=0$, $p=0$, $\gamma=0$, $\alpha\sim -c$ as $c\to
-\infty$.  The ground state is a nondegenerate state with trimer
character.
\item (0,1): $C(0,1)=0$, $\alpha\sim -c$ and $\gamma\to 0$ as $c\to
-\infty$. Asymptotic trimer character. Similar to the ground state but
it has a degenerate partner and $p\ne 0$.
\item $(0,n_2^0>1)$: $C(0,n_2^0)=0$, $\gamma\to (-2/3 n_2^0+1)\pi$ and
$\alpha\sim -c/2$ as $c\to -\infty$. Dimer character.
\item  $(1,n_1^0\ge 1)$: $-6\le C(1,n_2^0)<-4$,
$\gamma\to-2\pi/3(n_2^0-1)$, $\alpha\sim -c/2$ as $c\to-\infty$.  Also
dimer character, but it takes a stronger interaction to achieve in
comparison to the previous group.
\item $(n_1^0>1,n_2^0\ge n_1^0)$: Real $k_i$ for all $c$,
$\delta_j(\pm\infty)=2\pi n_j^0\pm1$. The energy tends to a constant
value as $|c|\to\infty$.
\end{itemize}

        It is hoped that the root structure found and the techniques
developed for its study will be useful for the examination of variants
of the model involving different interactions and/or an arbitrary
number of particles.  Since all eigenstates can be obtained for a
given $c$, one of the possible applications of the present work is the
simulation of time dependent rearrangement processes using a
discretized basis.  In this context it may serve as a reference exact
model to compare with approximate wave function propagation methods
based on periodic boundary conditions \cite{TD}. The model may also
permit an explicit comparison between the classical concepts of
``bound pair'' and ``free'' particles and their collisional
rearrangements with their quantum counterparts.  The origin of the
difficulties in the quantum case is that the Hamiltonian for the 3 (or
$N$) body system does not commute with the Hamiltonian of a pair, so
that using the concept of the bound pair in a gas (or in a box) is a
delicate matter \cite{RaSn}.

\acknowledgments{This work was supported in part by the Natural
Sciences and Engineering Research Council of Canada.  J. G. M.
acknowledges support by {\it Gobierno Aut\'onomo de Canarias (Spain)}
(Grant No. PI2/95) and {\it Ministerio de Educaci\'on y Ciencia
(Spain)} (PB 93-0578).  The authors thank Saman Alavi for suggesting
the use of a ternary phase diagram to represent the three particle
states.}

\begin{appendix}
\section{Normalization and potential energy}

        The expression (\ref{Bet}) for the wave function is not
normalized.  For the analysis of the energy spectrum this does not
have any effect.  However the calculation of physical expectation
values, and in particular of the potential energy, requires a
normalized function.  This requires the evaluation of the inner
product $\langle \psi|\psi\rangle$, where $|\psi\rangle$ is the
unnormalized total wave function given explicitly by Eq. (\ref{Bet})
in the region $R_{123}$. The contribution of the six regions is
identical so
\begin{equation}
\langle \psi|\psi\rangle=6
\int_0^1 dx_3\int_0^{x_3} dx_2 \int_0^{x_2} dx_1 
|\psi(k_1,k_2,k_3;x_1,x_2,x_3)|^2
\end{equation}
This integral is decomposed into 36 terms with integrals of the
general form
\begin{equation}
\int_0^1 dx_3 \int_0^{x_3} dx_2\int_0^{x_2} dx_1 
e^{i(\alpha_1 x_1+\alpha_2 x_1+\alpha_3 x_1)}
\end{equation}
where the $\alpha$'s are combinations of momenta of the form
$-k_j^*+k_i$, $(i,j=1,2,3)$.  These integrals can of course be solved
explicitly but the result is so lengthy that it is not reported here.
In order to handle all terms efficiently, it is useful to classify the
possible types of integrals.  There are three cases: ($\alpha_j=0,\,\,
j=1,2,3$), ($\alpha_j=0,\,\,\alpha_k+\alpha_i=0$), and ($\alpha_j\ne
0, j=1,2,3$). In all cases the sum of the $\alpha$'s is zero, $\sum_j
\alpha_j=p-p=0$.

        Once $\langle \psi|\psi\rangle$ is obtained, the
(dimensionless) potential energy involves a simpler integration
because of the delta functions, namely
\begin{equation}
\langle V\rangle=\frac{6c}{\langle\psi|\psi\rangle}
\int_0^1 dx_3\int_0^{x_3} dx_1 |\psi(x_1,x_1,x_3)|^2\,.
\end{equation}
Again, all integrals involved can be carried out.        
\end{appendix}

\vskip 12pt

\newpage

FIGURE CAPTIONS
{\bf Fig. 1} $\langle E\rangle$ vs $c$ for ($n^0,n^0$). From bottom to
top, $n^0=0,1,2,3$.  The lines for $n^0=0$ and 1 extend to $-\infty$
as $c\to-\infty$.

{\bf Fig. 2} $\langle V\rangle$ vs $c$ for ($n^0,n^0$).  From bottom
to top, $n^0=0,1,2,3$.  The lines for $n^0=0$ and 1 extend to
$-\infty$ as $c\to-\infty$.

{\bf Fig. 3} $\delta_1$ vs $c$. The numbers close to each line
correspond to $n_1^0$ and $n_2^0$. The figure for $\delta_2$ is
identical by interchanging $n_1^0$ and $n_2^0$. The lines are only
drawn up to the critical values of $c$ where $\delta_1=0$ [for (1,1),
(1,2) and (1,3)], or $\delta_2=0$ [for (2,1) and (3,1)]. For more
negative values of $c$, the $\alpha$-$\gamma$ parameterization is
used, see Figs. 5 and 6.

{\bf Fig. 4} $\delta$ vs $c$ for ($n^0,n^0$). 
From bottom to top, $n^0=0,1,2,3$.

{\bf Fig. 5} $\alpha$ vs c. Solid lines: $n_1^0=0$;
Dashed lines: $n_1^1=1$. The number close to the lines is $n_2^0$.

{\bf Fig. 6} $\gamma$ vs c. Solid lines:  $n_1^0=0$;
Dashed lines: $n_1^0=1$. The number close to the lines is $n_2^0$.

{\bf Fig. 7} Contour plot of the probability density of the state
(3,3) at $c=1$. The interpretation of a ``ternary phase'' type of
diagram is explained in the text.

{\bf Fig. 8} Contour plot of the probability density of the state
(0,0) at $c=-9$.

{\bf Fig. 9} Contour plot for the probability density of the state
(0,2) at $c=-9$.
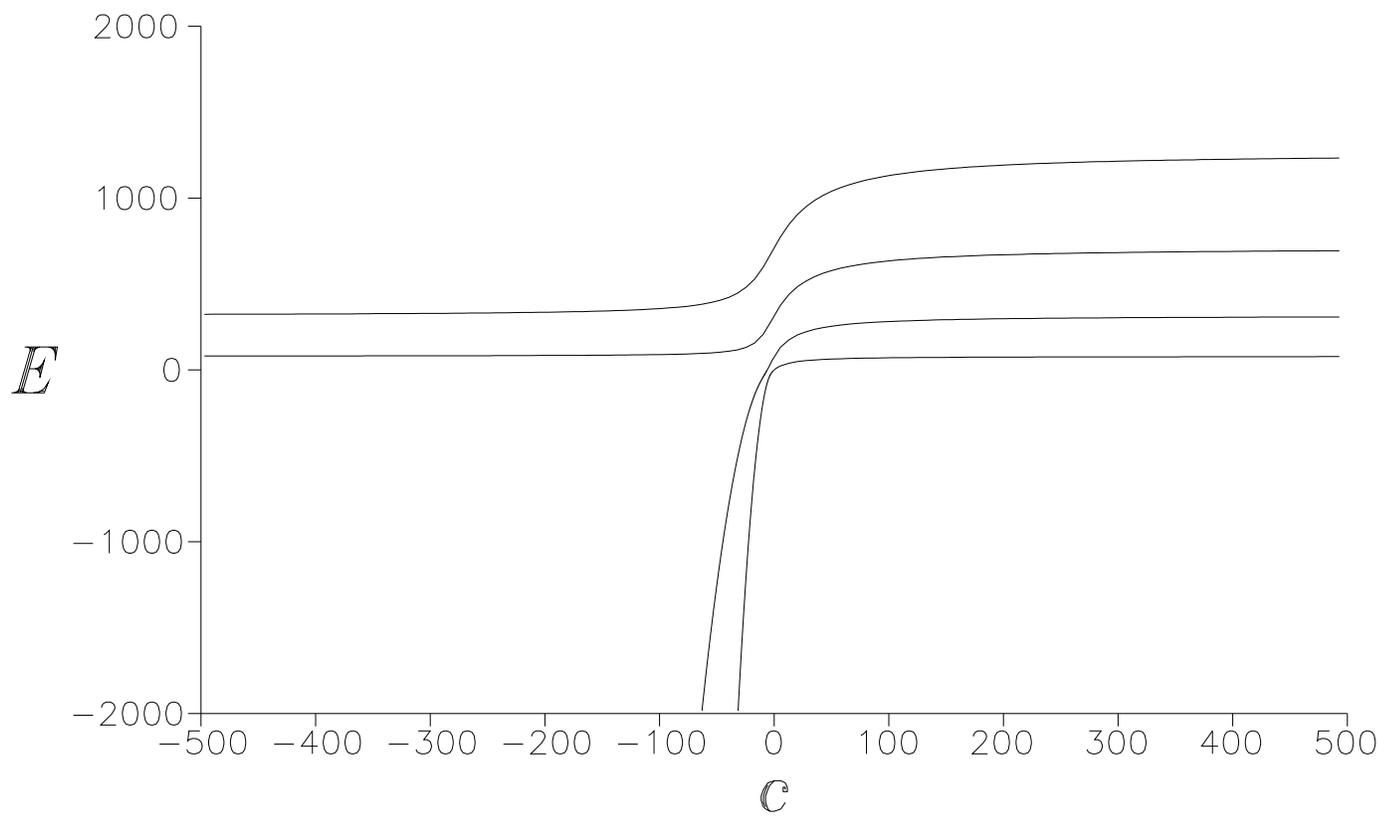
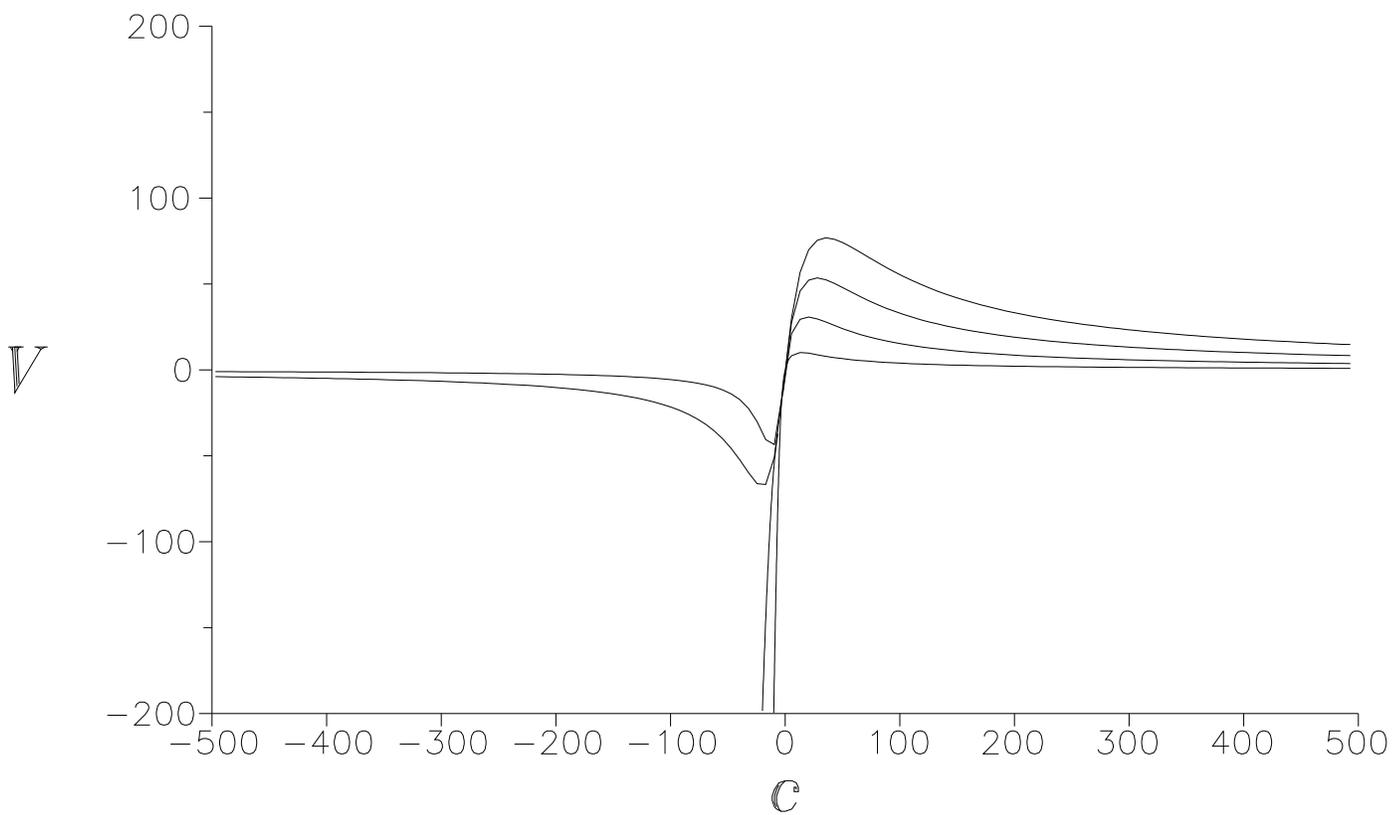
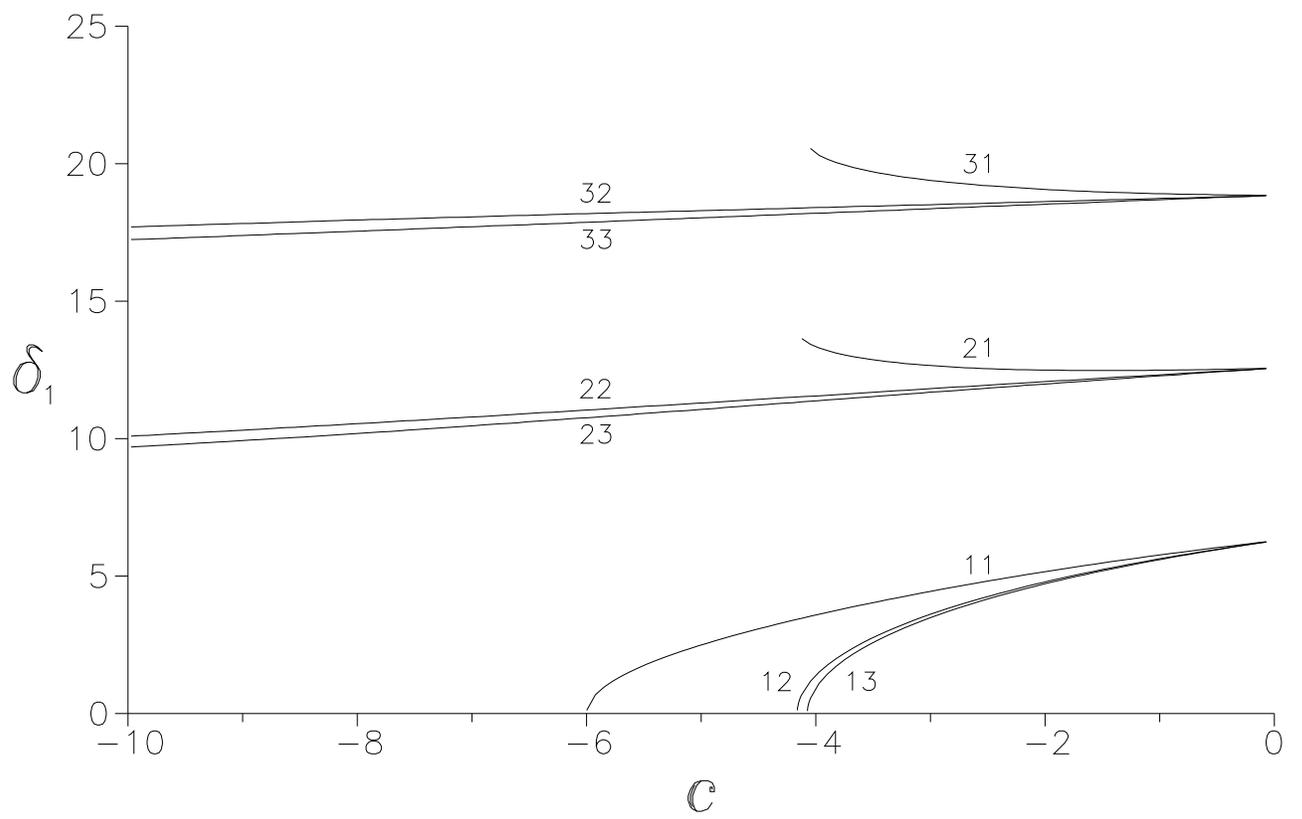
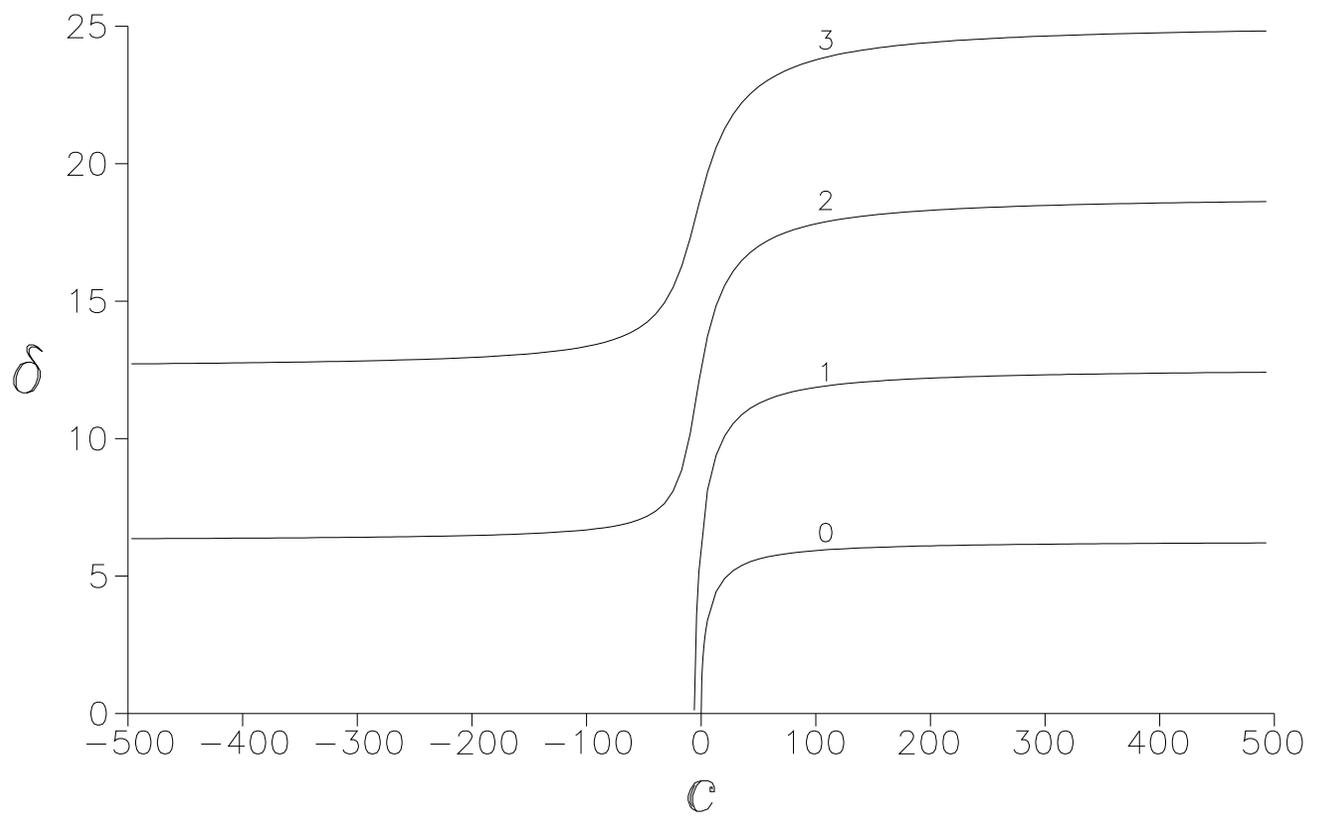
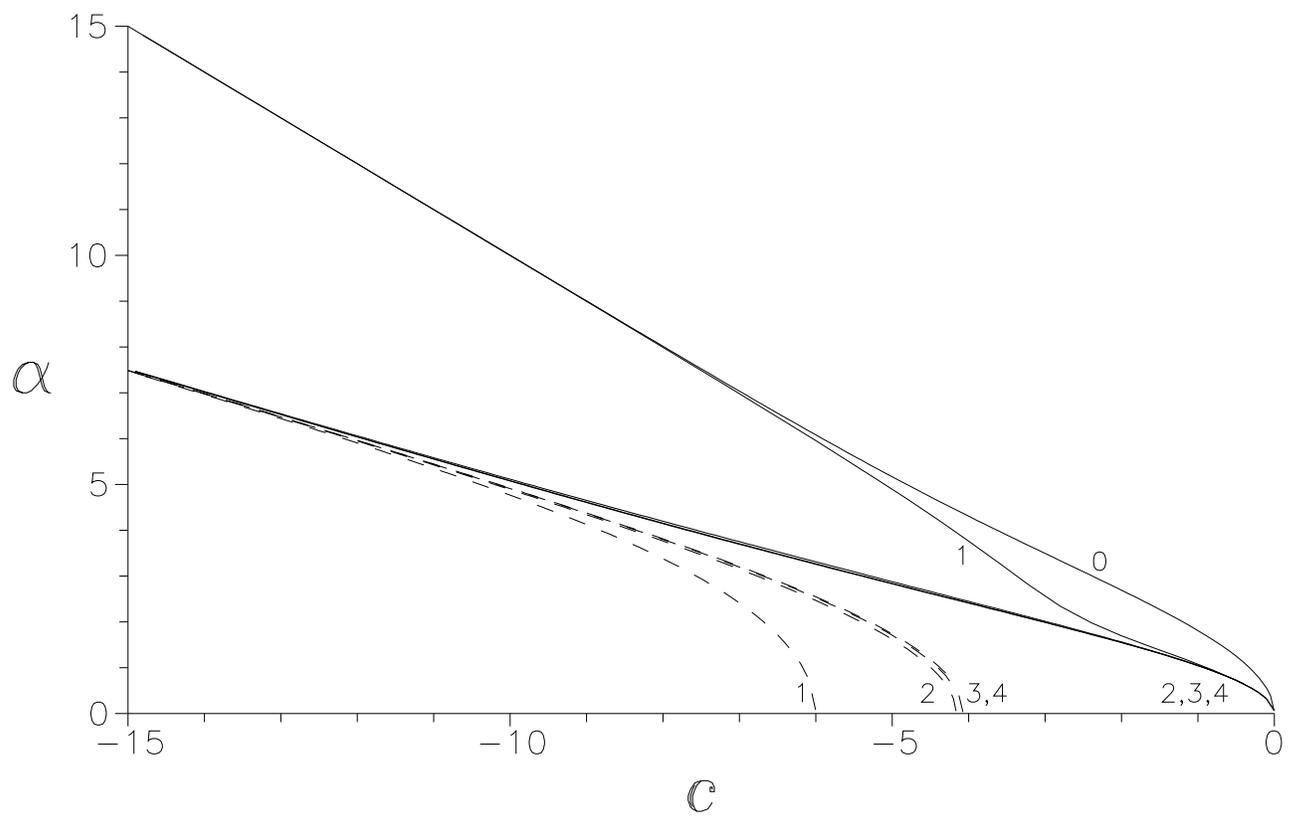
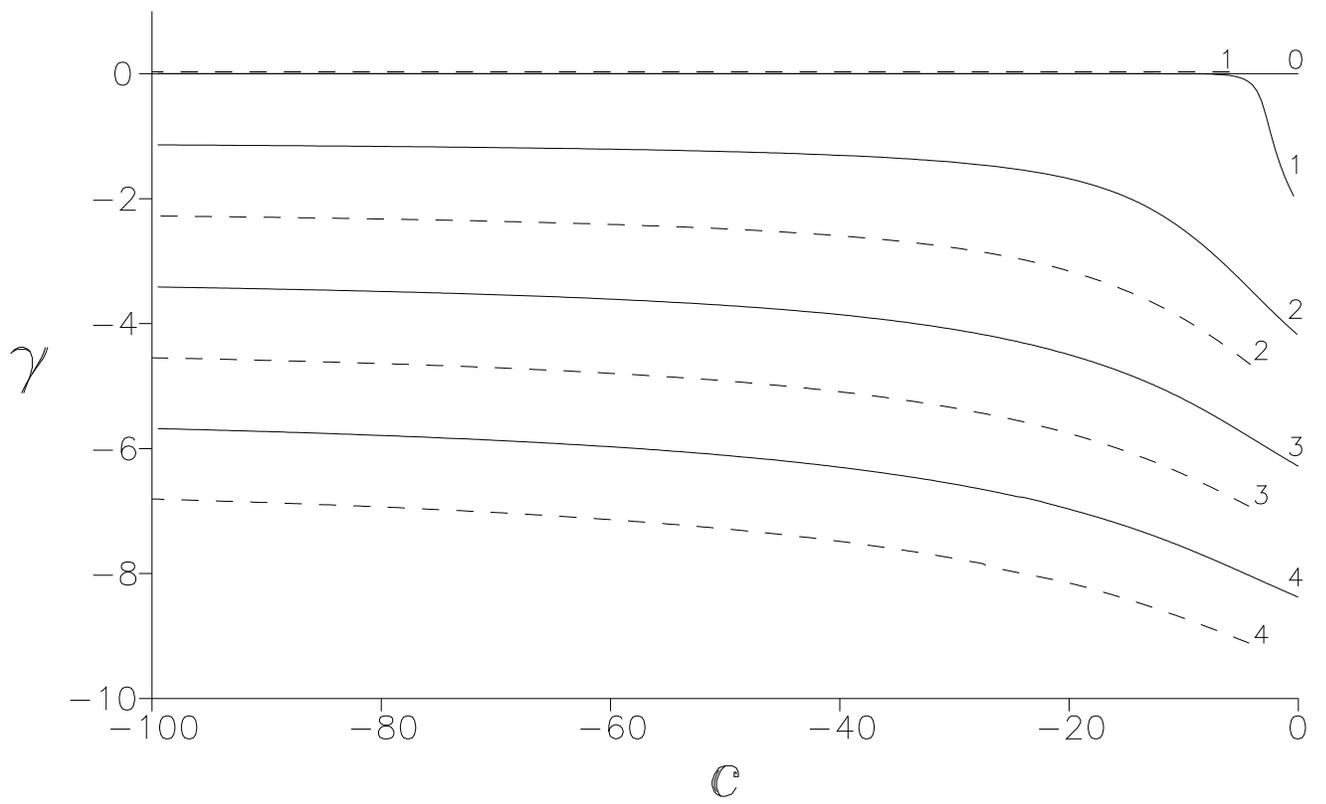
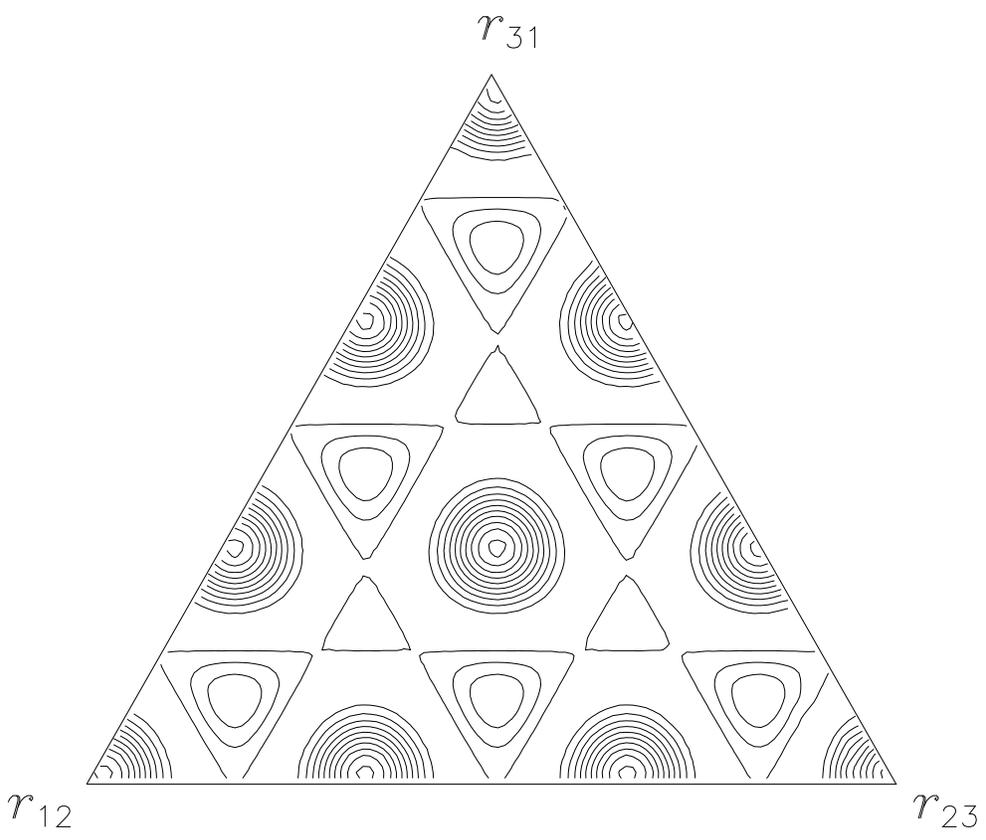
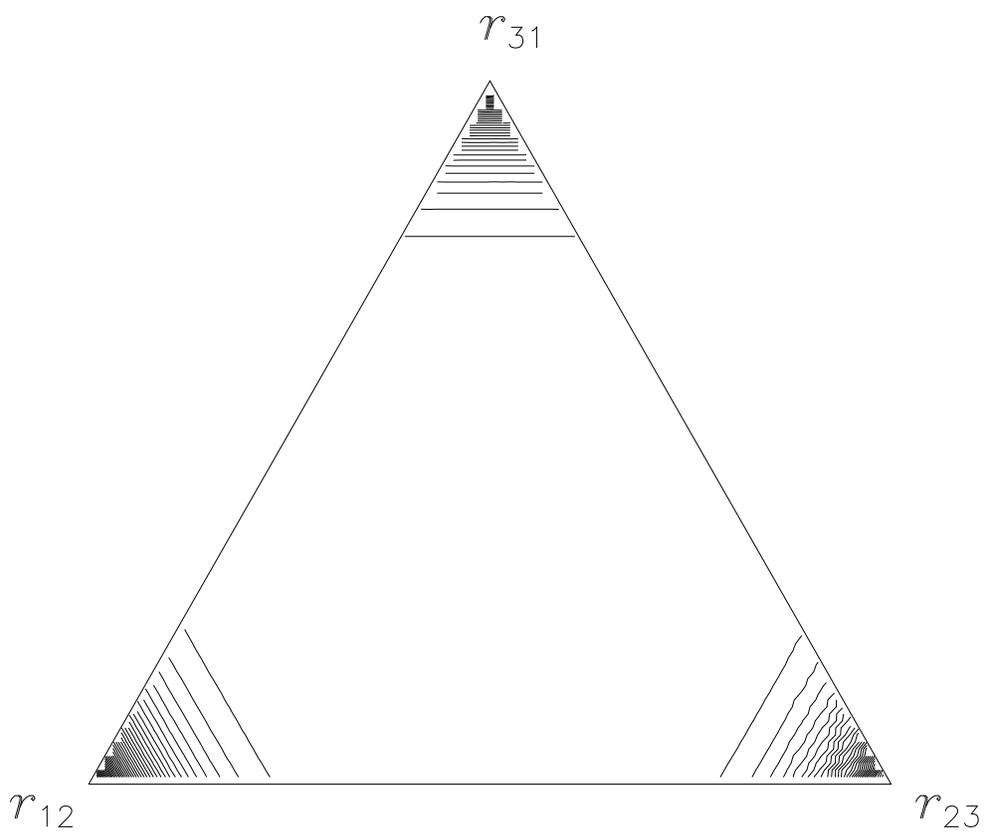
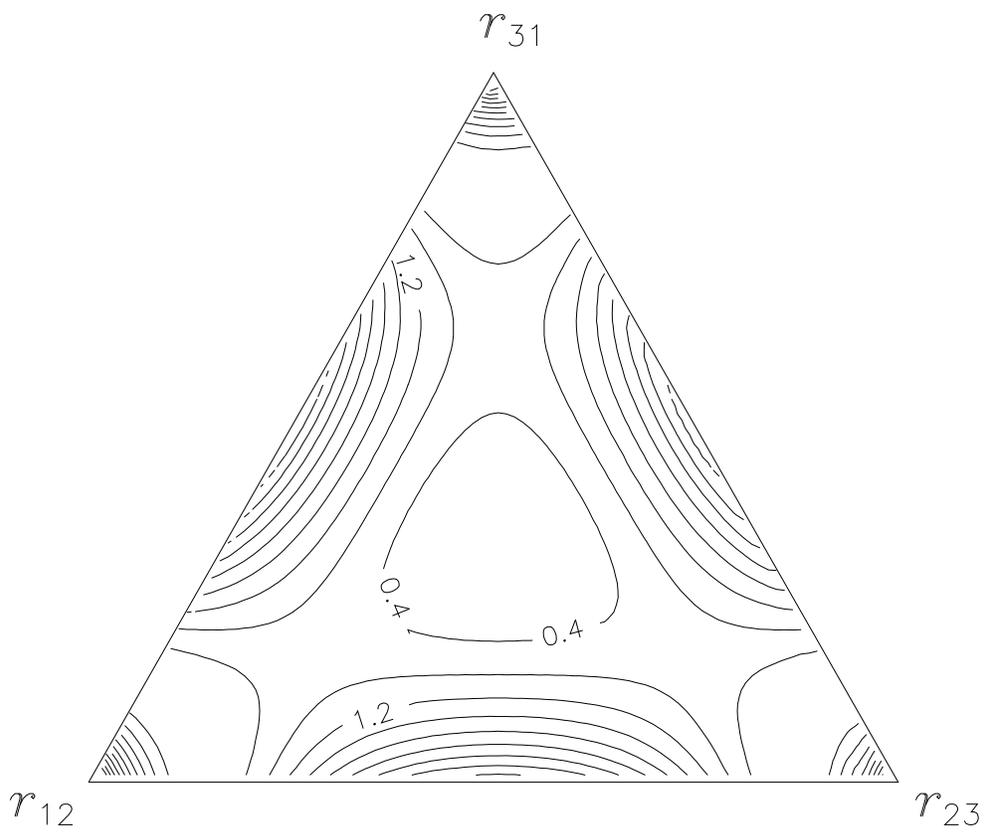
\end{document}